\documentclass[aps,prd,eqsecnum,onecolumn,showpacs,nofootinbib,preprintnumbers,a4paper]{revtex4}
\usepackage{amsmath}
\usepackage{amssymb}
\usepackage{amsfonts}
\usepackage{latexsym}
\usepackage{graphicx}
\usepackage{dcolumn}
\usepackage{bm}
\usepackage{empheq}
\usepackage{color}
\usepackage{times}  
\newcommand{\eq}[1]{\begin{equation}#1\end{equation}}

\newcommand{\ea}[1]{\begin{equation}\begin{aligned}#1\end{aligned}\end{equation}}

\newcommand{\lrb}[1]{\left( #1 \right)}  

\definecolor{gxhighlight}{rgb}{1,1,0.4}

\begin{document}

\preprint{CAS-KITPC/ITP-187}

\title{On non-linear CMB temperature anisotropy from gravitational perturbations}

\author{Xian Gao}%
    \email{gaoxian@itp.ac.cn}
    \affiliation{%
    \bigskip
        Kavli Institute for Theoretical Physics China,\\
        Key Laboratory of Frontiers in Theoretical Physics,
        Institute of Theoretical Physics, Chinese Academy of Sciences,\\
        No.55, ZhongGuanCun East Road, HaiDian District, Beijing, 100190, China
        }%


\pacs{Valid PACS appear here}
\keywords{}

\begin{abstract}

Non-linear CMB temperature anisotropies up to the third-order on
large scales are calculated. On large scales and in the Sachs-Wolfe
limit, we give the explicit expression for the observed temperature
anisotropy in terms of the primordial curvature perturbation up to
third-order. We derived the final bispectrum and trispectrum of
anisotropies and the corresponding non-linear parameters, in which
the contributions to the observed non-Gaussianity from primordial
perturbations and from the non-linear mapping from primordial
curvature perturbation to the temperature anisotropy are
transparently separated.

\end{abstract}

\maketitle%

\section{Introduction}

In the past few years, extensive attention has been attracted
 to the investigation of cosmological perturbations \cite{Lyth:1998xn} beyond
the linear order. The importance of studying  non-linear
perturbations comes two aspects. Firstly, forthcoming experiments of
Cosmic Microwave Background (CMB) and Large-scale Structure (LSS)
will be able to detect the non-linear structures in these
perturbations. The observational detection of the non-linearities
through the statistical non-Gaussianity (see \cite{Bartolo:2004if}
for a comprehensive review and \cite{Komatsu:2010hc} for a review of
recent observational progress ) of perturbations has become one the
the primary targets of the cosmology. On the other hand,
non-linearities, which encode the \emph{interactions} in the early
universe, would definitely bring us new understandings of both the
early universe and the fundamental physics.

A large amount of efforts have been devoted to the calculation of
the statistics of curvature perturbation $\zeta$, like primordial
bispectrum and trispectrum on large scales, pioneered by Maldacena
\cite{Maldacena:2002vr}{\footnote{See e.g. \cite{Koyama:2010xj} for
a recent short review of methods and techniques in calculating and
analyzing primordial non-Gaussianities from inflationary models and
references therein.}}. However, these are not the  observed
non-Gaussianities of (e.g.) CMB temperature anisotropies
$\frac{\Delta T}{T}$. Conventionally, one may use the linear order
relation $\frac{\Delta T}{T} = -\frac{1}{3}\Phi = -\frac{1}{5}\zeta$
to evaluate the angular bispectrum or trispectrum of CMB, assuming
the contributions from the second-order or secondary effects are
negligible comparing to the primordial ones
\cite{Komatsu:2010fb,Komatsu:2010hc,Komatsu:2001rj}. However, in
light of increasingly precise observations, a full treatment of the
higher-order radiation transfer functions of the CMB anisotropies is
needed, which will allow us to make definite prediction of CMB
non-Gaussianities (see \cite{Bartolo:2010qu} for a recent review for
non-Gaussianity on the CMB).

The research on non-linear temperature anisotropy due to
gravitational perturbations was pioneered by
\cite{Pyne:1995bs,Mollerach:1997up,Matarrese:1997ay}, in which the
second-order generalization of Sachs-Wolfe (SW) effect and
Integrated Sachs-Wolfe (ISW) effect were derived. These results were
extended in
\cite{Bartolo:2005kv,Bartolo:2004ty,Tomita:2005et,Tomita:2007db,Boubekeur:2009uk}
where the second-order radiation transfer function on large scales
was calculated, and in \cite{D'Amico:2007iw} where the general
expression for anisotropy due to gravitational perturbations up to
the third-order and in \cite{Bartolo:2005fp} where an elegant and
non-perturbative non-linear anisotropy were got. The non-linear
anisotropies have also been analyzed in
\cite{Maartens:1998xg,Dunsby:1997xy}  base on the covariant approach
to cosmological perturbations (see \cite{Langlois:2010vx} for a
recent review). Various secondary contributions to the
non-Gaussianities have also been extensively studied, including the
weak gravitational lensing and its correlation with ISW effect
\cite{Seljak:1998nu,Goldberg:1999xm,Serra:2008wc,Hanson:2009kg,Mangilli:2009dr,Creminelli:2004pv,Cooray:2002ee},
which is expected to be the dominant contamination of
$f_{\textrm{NL}}^{\textrm{local}}$, correlation between lensing and
Sunyaev-Zel'dovich (SZ) effect \cite{Goldberg:1999xm}, inhomogeneous
recombination \cite{Khatri:2008kb,Senatore:2008vi,Senatore:2008wk},
small-scale dark matter clustering
\cite{Bartolo:2008sg,Pitrou:2008ak}. A systematic treatment of the
transfer function on all scales, which involves solving  the full
Boltzmann equations through the recombination phase and then from
the surface of last-scattering until today, has been performed in
\cite{Bartolo:2006cu,Bartolo:2006fj,Nitta:2009jp,Pitrou:2008ak,Pitrou:2008hy,Pitrou:2008ut,Pitrou:2010sn,Beneke:2010eg}
at second-order.

In this note, we calculate the CMB temperature anisotropy up to the
third-order in primordial curvature perturbation $\zeta$, which can
be viewed as non-linear generalization of linear-order relation
$\Delta T/T = -\zeta/5$. We follow the same strategy in
\cite{Pyne:1995bs,Mollerach:1997up,Matarrese:1997ay,D'Amico:2007iw,Boubekeur:2009uk}.
First, we calculate the gravitational redshift of a given photon
from the emission
 surface to today's observer, which will give the
observed anisotropy in terms of metric perturbations $\frac{\Delta
T}{T} = \frac{\Delta T}{T}[\Phi,\Psi,\cdots]$. Then by using the
conservation of curvature perturbation $\zeta$, in the large-scale
limit, we determine the initial conditions for the metric
perturbations in matter-dominated era, more precisely,  the values
$\Phi_e = \Phi_e[\zeta]$ and $\Psi_e= \Psi_e[\zeta]$ on the emission
surface. Combining these two procedures will give the final
non-linear mapping from $\zeta$ to $\frac{\Delta T}{T}$:
$\frac{\Delta T}{T}=\frac{\Delta T}{T}[\zeta]$. Together with
previous results of primordial non-Gaussianities of $\zeta$  got in
the literature, our formalism is ready to make prediction of the
final observed CMB non-Gaussianity. Obviously, since we match the
metric perturbations in matter era directly to those on the emission
surface after last-scattering, our formalism includes only the
gravitational redshift of photons, without considering the dynamics
of photon-baryon plasma. Thus our result is valid only for
large-scale anisotropies, which enter the horizon after decoupling.


This note is organized as follows. In the following section, we
describe the temperature anisotropy induced by the gravitational
perturbations from the emission surface to the observers. We give
the general expression for the temperature anisotropy up to the
third-order in metric perturbations. In the next section, by using
the conserved curvature perturbation and taking the large-scale
limit, we determine the initial conditions for metric perturbations
in matter-dominated era. Then we give the non-linear mapping from
primordial curvature perturbation to the temperature anisotropy, in
the Sachs-Wolfe limit. Finally we give a short conclusion.


\section{Formalism}

After decoupling, the CMB photon density remains as Planck
distribution, which is determined by a single parameter --- the
photon temperature. The temperature shift of a Planck distribution
of photons is exactly proportional to the energy shift of any given
photon, i.e. $T/\omega = const$, if there is no collision, which
implies
    \eq{{\label{et_shift}}
        \frac{T_f}{T_i} = \frac{\omega_f}{\omega_i} \,,
    }
where $T_f$ and $T_i$ are the final and initial temperature of the
Planck distribution respectively, $\omega_f$ and $\omega_i$ are the
final and initial energy of a given photon respectively.
(\ref{et_shift}) is exact, which implies that in order to get the
change in the temperature, we need to evaluate the change in the
energy of a  (any) given photon, which in our question is nothing
but its gravitational redshift. Thus, the question becomes to follow
the geodesic equation of a given photon from the last-scattering
surface to us, taking the inhomogeneous spacetime background into
account.

We work in the generalized Poisson gauge
    \ea{{\label{pert_metric}}
        ds^2 = a^2 g_{\mu\nu} dx^{\mu}dx^{\nu}= a^2\lrb{ -e^{2\Phi} d\eta^2 + 2\sigma_i d\eta dx^i + e^{-2\Psi}e^{\gamma_{ij}}
        dx^idx^j } \,,
            }
where\footnote{Here and in what follows, two repeated lower or upper
spatial indices are contracted by $\delta_{ij}$} $\sigma_{i,i}=0$,
$\gamma_{ij}$ is symmetric and satisfies
$\gamma_{ij,i}=\gamma_{ii}=0$ (thus $\det e^{\gamma_{ij}}=1$), $a$
is the scale factor. The energy of a given photon with physical
momentum $P^{\mu}$ measured by an observer with 4-velocity
$u^{\mu}\equiv v^{\mu}/a$ (normalized as
$a^2g_{\mu\nu}u^{\mu}u^{\nu}=g_{\mu\nu}v^{\mu}v^{\nu}=-1$)
is{\footnote{As is well-known, metric $\tilde{g}_{\mu\nu}\equiv
e^{2\alpha}g_{\mu\nu}$ and $g_{\mu\nu}$ have the same null
geodesics, but parameterized by different affine parameters
$\tilde{\lambda}$ and $\lambda$ with relation $d\lambda =
e^{2\alpha}d\tilde{\lambda}$. Thus the energy in metric
$\tilde{g}_{\mu\nu}$ can be expressed as $\omega\equiv
-\tilde{g}_{\mu\nu}u^{\mu} \tilde{p}^{\nu} =
-g_{\mu\nu}u^{\mu}p^{\nu}$, where $\tilde{p}^{\mu}=
dx^{\mu}/d\tilde{\lambda}$ and $p^{\mu}= dx^{\mu}/d\lambda$ are
momentum in metric $\tilde{g}_{\mu\nu}$ and $g_{\mu\nu}$
respectively.}}
    \eq{{\label{eq:omega_def}}
        \omega = -a^2 g_{\mu\nu} u^{\mu} P^{\nu} = -g_{\mu\nu} u^{\mu} p^{\nu}
        \,,
    }
where $p^{\mu}$ is the momentum associated with the conformal metric
$g_{\mu\nu}$ (note $p^{\mu} = P^{\mu}/a^2$). Under the perturbed
metric (\ref{pert_metric}) and using normalization for $u^{\mu}$ and
$p^{\mu}$ to express $u^0$ and $p^0$ in terms of $u^i$ and $p^i$,
$\omega$ in terms of all relevant components takes the form:
    \eq{{\label{omega_ui_pi}}
        \omega =
        \frac{1}{a} \left( \sqrt{\tilde{g}_{ij}p^{i}p^{j}}\sqrt{\tilde{g}_{ij}v^{i}v^{j}+1}-\tilde{g}_{ij}v^{i}p^{j} \right),
    }
with
    \eq{{\label{eq:tilde_g_def}}
        \tilde{g}_{ij}\equiv e^{-2\Psi}e^{\gamma_{ij}}+e^{-2\Phi}\sigma_{i}\sigma_{j}.
    }
From (\ref{omega_ui_pi}), on the background level
$\bar{\omega}=|\bar{p}^i|/a = \bar{p}^0/a$, which we will normalize
by setting $\bar{p}^0=1$ in the following.
 The explicit expression for $u^i$ (or $v^i$) requires details of the
dynamics during recombination.

Assuming the ``intrinsic" photon temperature anisotropy at emission
point $\bm{x}_e$ in the direction ${\bm{n}}_e$ takes the form
$T\left(\eta_{e},x_{e}^{i};n_{e}^{i}\right) =
\bar{T}\left(\eta_{e},\bar{x}_{e}^{i};n^{i}\right)e^{\tau\left(\eta_{e},x_{e}^{i};n_{e}^{i}\right)}$,
where $\eta_e$ is the constant conformal time of emission, e.g. when
the last-scattering takes place. The temperature measured by an
observer at $\bm{x}_o$ and in the direction $\bm{n}$ is given by
    \eq{{\label{anisotropy_general}}
        \frac{\Delta T}{T}\left(x_{o}^{i},n^{i}\right) \equiv
        \frac{T\left(x_{o}^{i},n^{i}\right)-\bar{T}\left(x_{o}^{i},n^{i}\right)}{\bar{T}\left(x_{o}^{i},n^{i}\right)}
        =\frac{a_{o} \omega_{o}\left(\eta_{o},x_{o}^{i};p_{o}^{\mu}\right)}{a_{e} \omega_{e}\left(\eta_{e},x_{e}^{i};p_{e}^{\mu}\right)}e^{\tau\left(\eta_{e},x_{e}^{i};n_{e}^{i}\right)}-1
        ,
    }
where we have used
$\bar{T}_{o}=\frac{\bar{\omega}_{o}}{\bar{\omega}_{e}}\bar{T}_{e}=\frac{a_{e}}{a_{o}}\bar{T}_{e}$.
From (\ref{anisotropy_general}) it is clear that the observed
anisotropy comes from two aspects: the intrinsic anisotropy $\tau$
on the emission surface which depends on the dynamics during
recombination and the gravity theory, the other is the gravitational
redshift-induced anisotropy from the emission surface to the
observer ${a_o\omega_o}/{(a_e\omega_e)}$, which is purely kinetic
and is independent of the theory of gravitation. The intrinsic
anisotropy $\tau$ is highly model-dependent and a full treatment
needs solving the set of Boltzmann equations up to the third-order
which is beyond the scope of this note. On super-Horizon (the sound
horizon on the last-scattering surface) scales where microscopic
physics is irrelevant, a simple and non-perturbative expression has
been got \cite{Bartolo:2005fp}: $\tau = -\frac{2}{3}\Phi$, which is
adequate for our purpose.

In the following, first we derive the gravitational redshift-induced
anisotropy up to the third-order in terms of metric perturbations in
(\ref{pert_metric}), then use the Einstein equation to determine the
initial condition at matter-dominated era, i.e. the values of metric
perturbations in terms of the conserved curvature perturbation
$\zeta$ on large scales.

\subsection{Perturbed photon energy}
 After
decoupling, the photons propagate freely and thus the physical
content of its Boltzmann equation is completely encoded in the
photon geodesic equation, which is much simpler to deal with. Null
geodesics in perturbed spacetime has been investigated long before
\cite{Pyne:1993np,Pyne:1995bs,Mollerach:1997up,Matarrese:1997ay,D'Amico:2007iw,Boubekeur:2009uk}.
Although the whole thing we need to do is simply to follow the
redshift of a photon, in a practical calculation, the complexities
arise from several aspects. First, the energy $\omega$ should be
evaluated at the ``real" emission point $x^{i}(\lambda_e)$ rather
than at the ``virtual image" at $\bar{x}^{i}(\bar{\lambda}_e)\equiv
(\lambda_e-\lambda_o)n^i$. Here $\lambda$ is the affine parameter
along the photon geodesics, $\lambda_e$ ($\lambda_o$) is the
corresponding values at emission surface (observer). Second, to
determine the real position of emission we need to follow the photon
geodesics, which we are able to solve only perturbatively.

The expansion of frequency $\omega$ around the background emission
point $x^i_o$ and the direction $n^i\equiv -\bar{p}^i$ is
straightforward
\cite{Pyne:1995bs,Mollerach:1997up,Matarrese:1997ay,Bartolo:2004ty,D'Amico:2007iw}.
Here we simply report the corresponding expansions of
 (\ref{omega_ui_pi}) up to the third-order in the  metric perturbations (\ref{pert_metric}), which involve the spatial components $v^i$ and
 $p^i$. At the linear-order:
    \eq{{\label{omega_1_bare}}
    a\omega_{\left(1\right)} =
    \left(v_{\left(1\right)}^{i}-p_{\left(1\right)}^{i}\right)n^{i}-\Psi+\frac{1}{2}\gamma_{ij}n^{i}n^{j}
    ,
    }
where throughout this note we take the expansion of variable $Q$ as
$Q = \bar{Q}+Q_{(1)}+ Q_{(2)}+Q_{(3)}+\cdots$. At second-order:
    \ea{{\label{omega_2_bare}}
        a\omega_{\left(2\right)} =&\left(v_{\left(2\right)}^{i}-p_{\left(2\right)}^{i}+x_{\left(1\right)}^{0}\dot{p}_{\left(1\right)}^{i}\right)n^{i}+\frac{1}{2}\Psi^{2}+\Psi n^{i}\left(p_{\left(1\right)}^{i}-2v_{\left(1\right)}^{i}\right)+\frac{1}{2}\left(p_{\left(1\right)}^{i}-v_{\left(1\right)}^{i}\right)^{2}-\frac{1}{2}\left(n^{i}p_{\left(1\right)}^{i}\right)^{2}\\&+\frac{1}{4}\left[\gamma_{ij}^{2}-2\left(\Psi-n^{k}p_{\left(1\right)}^{k}+\frac{1}{4}\gamma_{kl}n^{k}n^{l}\right)\gamma_{ij}\right]n^{i}n^{j}+\gamma_{ij}n^{i}\left(v_{\left(1\right)}^{j}-p_{\left(1\right)}^{j}\right)\\&+\left(x_{\left(1\right)}^{k}+x_{\left(1\right)}^{0}n^{k}\right)\partial_{k}\left(v_{\left(1\right)}^{i}n^{i}-\Psi+\frac{1}{2}\gamma_{ij}n^{i}n^{j}\right)+\frac{1}{2}\left(\sigma_{i}n^{i}\right)^{2}.
    }
and
    \ea{\label{omega_3_bare}
        a\omega_{\left(3\right)} =&n^{i}\Big[v_{\left(3\right)}^{i}-p_{\left(3\right)}^{i}+\left(x_{\left(1\right)}^{j}+x_{\left(1\right)}^{0}n^{j}\right)\left(\partial_{j}v_{\left(2\right)}^{i}+\frac{1}{2}\left(x_{\left(1\right)}^{k}+x_{\left(1\right)}^{0}n^{k}\right)\partial_{j}\partial_{k}v_{\left(1\right)}^{i}\right)+\left(x_{\left(2\right)}^{j}-x_{\left(1\right)}^{0}p_{\left(1\right)}^{i}\right)\partial_{j}v_{\left(1\right)}^{i}\\&+x_{\left(1\right)}^{0}\dot{p}_{\left(2\right)}^{i}+\left(x_{\left(2\right)}^{0}-x_{\left(1\right)}^{0}p_{\left(1\right)}^{0}\right)\left(n^{j}\partial_{j}v_{\left(1\right)}^{i}+\dot{p}_{\left(1\right)}^{i}\right)-\frac{1}{2}\left(x_{\left(1\right)}^{0}\right)^{2}\ddot{p}_{\left(1\right)}^{i}\Big]\\&+\left[\left(\Psi-n^{j}p_{\left(1\right)}^{j}+\frac{1}{2}\gamma_{kl}n^{k}n^{l}\right)n^{i}-\gamma_{ij}n^{j}+p_{\left(1\right)}^{i}-v_{\left(1\right)}^{i}\right]\left(p_{\left(2\right)}^{i}-x_{\left(1\right)}^{0}\dot{p}_{\left(1\right)}^{i}\right)\\&+\left(v_{\left(1\right)}^{i}-p_{\left(1\right)}^{i}+\gamma_{ij}n^{j}-2\Psi n^{i}\right)\left(v_{\left(2\right)}^{i}+\left(x_{\left(1\right)}^{j}+x_{\left(1\right)}^{0}n^{j}\right)\partial_{j}v_{\left(1\right)}^{i}\right)\\&-\left[\left(x_{\left(2\right)}^{k}+x_{\left(2\right)}^{0}n^{k}-x_{\left(1\right)}^{0}p_{\left(1\right)}^{0}n^{k}-x_{\left(1\right)}^{0}p_{\left(1\right)}^{k}\right)\partial_{k}+\frac{1}{2}\left(x_{\left(1\right)}^{k}+x_{\left(1\right)}^{0}n^{k}\right)\left(x_{\left(1\right)}^{l}+x_{\left(1\right)}^{0}n^{l}\right)\partial_{k}\partial_{l}\right]\left(\Psi-\frac{1}{2}\gamma_{ij}n^{i}n^{j}\right)\\&+\frac{1}{2}n^{k}p_{\left(1\right)}^{k}\left(p_{\left(1\right)}^{i}\right)^{2}-\frac{1}{2}\left(n^{i}p_{\left(1\right)}^{i}\right)^{3}-\Phi\left(\sigma_{i}n^{i}\right)^{2}-\frac{1}{6}\Psi^{3}-\Psi^{2}\left(\frac{1}{2}n^{i}p_{\left(1\right)}^{i}-\frac{1}{4}\gamma_{ij}n^{i}n^{j}-2v_{\left(1\right)}^{i}n^{i}\right)\\&+\Psi\Big[\frac{1}{2}\left(n^{i}p_{\left(1\right)}^{i}\right)^{2}-\frac{3}{2}\left(v_{\left(1\right)}^{i}\right)^{2}+\frac{1}{2}\left(4v_{\left(1\right)}^{i}-p_{\left(1\right)}^{i}\right)p_{\left(1\right)}^{i}\\&+\frac{1}{8}\left(\left(\gamma_{kl}n^{k}n^{l}-4n^{k}p_{\left(1\right)}^{k}\right)\gamma_{ij}-2\gamma_{ij}^{2}\right)n^{i}n^{j}+\gamma_{ij}n^{i}\left(p_{\left(1\right)}^{j}-2v_{\left(1\right)}^{j}\right)+\frac{1}{2}\left(\sigma_{i}n^{i}\right)^{2}\Big]\\&+\frac{1}{16}\left(\gamma_{ij}n^{i}n^{j}\right)\left[12\left(n^{k}p_{\left(1\right)}^{k}\right)^{2}-4\left(p_{\left(1\right)}^{k}\right)^{2}+8\gamma_{kl}n^{k}p_{\left(1\right)}^{l}+\left(\gamma_{kl}n^{k}n^{l}-6n^{k}p_{\left(1\right)}^{k}\right)\left(\gamma_{mn}n^{m}n^{n}\right)-2\gamma_{kl}^{2}n^{k}n^{l}\right]\\&+\frac{1}{2}\left[\left(p_{\left(1\right)}^{i}-2n^{k}p_{\left(1\right)}^{k}n^{i}\right)p_{\left(1\right)}^{j}+v_{\left(1\right)}^{i}\left(v_{\left(1\right)}^{j}-2p_{\left(1\right)}^{j}\right)\right]\gamma_{ij}+\frac{1}{4}\left(n^{k}p_{\left(1\right)}^{k}n^{j}+2v_{\left(1\right)}^{j}\right)\gamma_{ij}^{2}n^{i}+\frac{1}{12}\gamma_{ij}^{3}n^{i}n^{j}\\&+\left(x_{\left(1\right)}^{k}+x_{\left(1\right)}^{0}n^{k}\right)\Big\{\left[\Psi-\frac{1}{2}\gamma_{ij}n^{i}n^{j}+n^{i}\left(p_{\left(1\right)}^{i}-2v_{\left(1\right)}^{i}\right)\right]\partial_{k}\Psi-\frac{1}{8}\partial_{k}\left(\gamma_{ij}n^{i}n^{j}\right)^{2}\\&+\left(n^{i}\left(v_{\left(1\right)}^{j}-p_{\left(1\right)}^{j}\right)-\frac{1}{2}\left(\Psi-n^{l}p_{\left(1\right)}^{l}\right)n^{i}n^{j}\right)\partial_{k}\gamma_{ij}+\left(\frac{1}{2}\gamma_{il}\partial_{k}\gamma_{lj}+\sigma_{i}\partial_{k}\sigma_{j}\right)n^{i}n^{j}\Big\}\\&+\sigma_{i}\sigma_{j}n^{i}\left(v_{\left(1\right)}^{j}-p_{\left(1\right)}^{j}\right)+\frac{1}{4}\left(2n^{i}p_{\left(1\right)}^{i}-\gamma_{ij}n^{i}n^{j}\right)\left(\left(\sigma_{k}n^{k}\right)^{2}-\left(v_{\left(1\right)}^{k}\right)^{2}\right).
    }
In (\ref{omega_1_bare})-(\ref{omega_3_bare}), a dot denotes
derivative with respect to $\lambda$ and subscripts ``${}_{(i)}$"
denote the orders in metric perturbations and all quantities are
evaluated on the background emission point (virtual
image){\footnote{(\ref{omega_1_bare})-(\ref{omega_3_bare}) are
 essentially equal to (e.g.) eq.(2.17)-(2.19) in \cite{D'Amico:2007iw}. The
 differences come from 1) here we have replaced $v^0$ and $k^0$ in terms of $v^i$, $k^i$ and metric perturbations through
the constraints $v^2=-1$ and $k^2=0$, 2) we use $\Phi$ and $\Psi$
rather than $\phi$ and $\psi$ which are defined as $1+2\phi\equiv
e^{2\Phi}$, $1-2\psi\equiv e^{-2\Psi}$ and 3) at this point we have
not expanded $\Phi=\Phi_{(1)} + \Phi_{(2)}+\Phi_{(3)}$ etc.}}. In
deriving the above results, we have used the expansion of
$\lambda_e$ around its background value: $\lambda_e =
\bar{\lambda}_e + \lambda_{(1)} +\lambda_{(2)}$ where $\lambda_{(1)}
= -x_{\left(1\right)}^{0}\left(\bar{\lambda}_{e}\right) $ and
$\lambda_{(2)} =
-x_{\left(2\right)}^{0}\left(\bar{\lambda}_{e}\right)+x_{\left(1\right)}^{0}\left(\bar{\lambda}_{e}\right)p_{\left(1\right)}^{0}\left(\bar{\lambda}_{e}\right)
$. This can be got by perturbing $x^0(\lambda_e) =
\bar{x}^0(\bar{\lambda}_e)\equiv \eta_e$, which is the definition of
the emission surface as intersection of past light-cone of the
observer and the spatial hypersurface at constant $\eta_e$.

\subsection{Photon geodesics}

(\ref{omega_1_bare})-(\ref{omega_3_bare}) can be fully determined
when the perturbed photon geodesics is solved. Geodesic equation in
the conformal metric $g_{\mu\nu}$ is
    \eq{{\label{geo_eq}}
        \dot{p}^{\mu} +\Gamma^{\mu}_{\rho\sigma} p^{\rho}
        p^{\sigma} = 0\,,
    }
where $p^{\mu}=dx^{\mu}/d\lambda$, a dot denotes $d/d\lambda$,
$\Gamma^{\mu}_{\rho\sigma}$ is the connection associated with the
conformal metric $g_{\mu\nu}$.

Since in (\ref{omega_1_bare})-(\ref{omega_3_bare}) we have expressed
the photon energy in terms of spatial momentum $p^i$, it is adequate
to solve $p^i$ perturbatively
\cite{Pyne:1993np,Pyne:1995bs,Mollerach:1997up,Matarrese:1997ay,D'Amico:2007iw,Boubekeur:2009uk}.
In order to make (\ref{geo_eq}) a close set of equations for $x^i$,
we also need the expression for $x^0$ in terms of $x^{i}$. This can
be done by perturbing the constraint $p_{\mu}p^{\mu}=0$ around the
background geodesics, which yields (up to the second-order in metric
perturbation)
    \eq{{\label{p0_1_constr}}
        p^0_{(1)} = -n^{i}p_{\left(1\right)}^{i}- A \,,
        }
with
    \eq{
        A \equiv
        \Phi+\Psi+\sigma_{i}n^{i}-\frac{1}{2}\gamma_{ij}n^{i}n^{j}
        \,,
    }
and \ea{{\label{p0_2_constr}}
    p_{\left(2\right)}^{0} =&   -n^{i}p_{\left(2\right)}^{i}+\frac{1}{2}\left(\delta_{ij}-n^{i}n^{j}\right)p_{\left(1\right)}^{i}p_{\left(1\right)}^{j}+\left[\left(\Phi+\Psi+\frac{1}{2}\gamma_{jk}n^{j}n^{k}\right)n^{i}+\sigma_{i}-\gamma_{ij}n^{j}\right]p_{\left(1\right)}^{i}\\
 &   -x_{\left(1\right)}^{\mu}\partial_{\mu}A+\frac{1}{2}\left(\Phi+\Psi\right)^{2}+2\Phi\sigma_{i}n^{i}-\left(\frac{1}{2}\left(\Phi+\Psi\right)\gamma_{ij}-\frac{1}{2}\sigma_{i}\sigma_{j}-\frac{1}{4}\gamma_{ij}^{2}\right)n^{i}n^{j}-\frac{1}{8}\left(\gamma_{ij}n^{i}n^{j}\right)^{2}.
 }
Integration of (\ref{p0_1_constr}) and (\ref{p0_2_constr}) with
respect to $\lambda$ along the background geodesics
$\bar{x}^{\mu}(\lambda)$ will give $x^0_{(1)}$ and $x^0_{(2)}$ in
terms of $x^i_{(1)}$ and $x^i_{(2)}$ respectively.

Having deriving the general expressions for perturbed photon energy
(\ref{omega_1_bare})-(\ref{omega_3_bare}), in the following, we
restrict ourselves to the large-scale limit. This is mainly because
that in our formalism, to eventually determine the observed
anisotropy in terms of conserved primordial curvature perturbation
$\zeta$, we use the values of metric perturbations at
matter-dominated era as initial conditions for gravitational
redshift of photons rather than for the full Boltzmann equations.
Thus, our results are only valid for the large-scale perturbation
modes, which enter the horizon after decoupling and never affected
by the microphysics. In the following we neglect the vector and
tensor metric perturbations, not only because on large scales vector
and tensor modes are subdominant but also calculation involving
these modes up to the third-order is rather cumbersome. We also
assume the the observer is comoving with the emission point, i.e
$v^i=0$. Under these assumptions, a non-perturbative approach to the
null geodesics has also been developed in \cite{D'Amico:2007iw}.

Following the logic in
\cite{Pyne:1993np,Pyne:1995bs,Mollerach:1997up,Matarrese:1997ay,D'Amico:2007iw,Boubekeur:2009uk},
for spatially-flat FRW background, the set of perturbed geodesic
equations are $\dot{p}^i_{(n)}=\ddot{x}^i_{(n)} = f^i_{(n)}$, with
    \begin{eqnarray}
        f_{\left(1\right)}^{i} &= &
        -\Gamma_{\rho\sigma\left(1\right)}^{i}\bar{p}^{\rho}\bar{p}^{\sigma},\\
        f_{\left(2\right)}^{i}&=&-\Gamma_{\rho\sigma\left(2\right)}^{i}\bar{p}^{\rho}\bar{p}^{\sigma}-2\Gamma_{\rho\sigma\left(1\right)}^{i}\bar{p}^{\rho}\dot{x}_{\left(1\right)}^{\sigma}-x_{\left(1\right)}^{\lambda}\partial_{\lambda}\Gamma_{\rho\sigma\left(1\right)}^{i}\bar{p}^{\rho}\bar{p}^{\sigma},\\
        f_{\left(3\right)}^{i}&=&
        -\Gamma_{\rho\sigma\left(3\right)}^{i}\bar{p}^{\rho}\bar{p}^{\sigma}-2\Gamma_{\rho\sigma\left(1\right)}^{i}\bar{p}^{\rho}\dot{x}_{\left(2\right)}^{\sigma}-2\Gamma_{\rho\sigma\left(2\right)}^{i}\bar{p}^{\rho}\dot{x}_{\left(1\right)}^{\sigma}-\Gamma_{\rho\sigma\left(1\right)}^{i}\dot{x}_{\left(1\right)}^{\rho}\dot{x}_{\left(1\right)}^{\sigma}\nonumber\\
        &&-2x_{\left(1\right)}^{\lambda}\partial_{\lambda}\Gamma_{\rho\sigma\left(1\right)}^{i}\bar{p}^{\rho}\dot{x}_{\left(1\right)}^{\sigma}-\left(x_{\left(1\right)}^{\lambda}\partial_{\lambda}\Gamma_{\rho\sigma\left(2\right)}^{i}+x_{\left(2\right)}^{\lambda}\partial_{\lambda}\Gamma_{\rho\sigma\left(1\right)}^{i}+\frac{1}{2}x_{\left(1\right)}^{\lambda}x_{\left(1\right)}^{\tau}\partial_{\lambda}\partial_{\tau}\Gamma_{\rho\sigma\left(1\right)}^{i}\right)\bar{p}^{\rho}\bar{p}^{\sigma}.
    \end{eqnarray}
The perturbed Christofel symbol can be read from
(\ref{app_connection_ls}). After some manipulations, we can solve,
at first-order in $\Phi$ and $\Psi$,
    \begin{eqnarray}
    p_{\left(1\right)}^{i} & = & -2n^{i}\Psi-I_{1}^{i}, \label{p^i_1}\\
    p_{\left(1\right)}^{0} & = & -2\Phi+I_{1},
    \end{eqnarray}
with
    \eq{
    I_{1}^{i} =
    \int_{\lambda_{o}}^{\lambda}d\tilde{\lambda}\partial_{i}A,\qquad
    I_{1} =
    \int_{\lambda_{o}}^{\lambda}d\tilde{\lambda}A',\qquad\textrm{where}~ A\equiv
    \Phi+\Psi.
    }
Here and in the following we will frequently use the trick: (e.g.)
$\dot{\Psi} = \left(\Psi'-n^{i}\partial_{i}\Psi \right)+\cdots$. At
second-order \cite{Pyne:1995bs,Mollerach:1997up,Matarrese:1997ay}:
    \begin{eqnarray}
    p_{\left(2\right)}^{i} & = & -2n^{i}\left(x_{\left(1\right)}^{\mu}\partial_{\mu}\Psi+\Psi^{2}\right)-2\Psi I_{1}^{i}+I_{2}^{i},\\
    p_{\left(2\right)}^{0} & = & 2\Phi^{2}-2x_{\left(1\right)}^{\mu}\partial_{\mu}\Phi-2I_{1}\Phi+I_{1}^{2}+I_{2},
    \end{eqnarray}
with
\begin{eqnarray}
    I_{2}^{i} & \equiv & \int_{\lambda_{o}}^{\lambda}d\tilde{\lambda}\left[2\left(\Phi-I_{1}\right)-x_{\left(1\right)}^{\mu}\partial_{\mu}\right]\partial_{i}A,\\
    I_{2} & \equiv &
    \int_{\lambda_{o}}^{\lambda}d\tilde{\lambda}\left(x_{\left(1\right)}^{\mu}\partial_{\mu}A'-2\Phi
    A'\right),\end{eqnarray}
At the third-order \cite{D'Amico:2007iw},
    \begin{eqnarray}
    p_{\left(3\right)}^{i} & = & -\frac{4}{3}n^{i}\Psi^{3}-2n^{i}\left(x_{\left(2\right)}^{\lambda}\partial_{\lambda}\Psi+\frac{1}{2}x_{\left(1\right)}^{\lambda}x_{\left(1\right)}^{\tau}\partial_{\lambda}\partial_{\tau}\Psi+2\Psi x_{\left(1\right)}^{\mu}\partial_{\mu}\Psi\right) {\nonumber}\\
    &  & -2\Psi^{2}I_{1}^{i}+2\Psi I_{2}^{i}-2I_{1}^{i}x_{\left(1\right)}^{\lambda}\partial_{\lambda}\Psi+I_{3}^{i},\label{p^i_3}\end{eqnarray}
with
    \begin{eqnarray}
    I_{3}^{i} & = & \int_{\lambda_{o}}^{\lambda}d\tilde{\lambda}\Big[2\left(\Phi-\Psi-I_{1}\right)x_{\left(1\right)}^{\mu}\partial_{\mu}\partial_{i}A-x_{\left(2\right)}^{\mu}\partial_{\mu}\partial_{i}A-\frac{1}{2}x_{\left(1\right)}^{\lambda}x_{\left(1\right)}^{\tau}\partial_{\lambda}\partial_{\tau}\partial_{i}A {\nonumber}\\
    &  & -\left(2\left(\Phi^{2}-2\Psi\Phi\right)+3I_{1}^{2}+2I_{2}+4\left(\Psi-\Phi\right)I_{1}-2x_{\left(1\right)}^{\mu}\partial_{\mu}\Phi\right)\partial_{i}A\Big].\end{eqnarray}

Finally, after plugging (\ref{p^i_1})-(\ref{p^i_3}) into
(\ref{omega_1_bare})-(\ref{omega_3_bare}), the perturbed photon
frequency up to the third-order in $\Phi$ and $\Psi$ is given by
    \eq{{\label{aomega_1}}
        a\omega_{\left(1\right)} = -\Phi+I_{1} ,
    }
    \eq{{\label{aomega_2}}
        a\omega_{\left(2\right)} = \frac{1}{2}\Phi^{2}+I_{2}-\Phi
        I_{1}+I_{1}^{2}-x_{\left(1\right)}^{0}A'-\left(x_{\left(1\right)}^{i}+x_{\left(1\right)}^{0}n^{i}\right)\partial_{i}\Phi
        ,
    }
and
    \ea{{\label{aomega_3}}
        a\omega_{\left(3\right)} =&-\frac{\Phi^{3}}{6}+\frac{\Phi^{2} {I}_{1}}{2}+I_{1}^{3}+I_{3}+I_{1}\left(2I_{2}-x_{1}^{0}A'\right)+\Phi\left(x_{\left(1\right)}^{0}A'-I_{1}^{2}-I_{2}\right)-x_{\left(2\right)}^{0}A'-\frac{1}{2}\left(x_{\left(1\right)}^{0}\right)^{2}A''\\&-\partial_{i}\Phi\left[n^{i}\left(\left(A+\Psi-2x_{\left(1\right)}^{0}\right)x_{\left(1\right)}^{0}+x_{\left(2\right)}^{0}\right)+x_{\left(2\right)}^{i}+x_{\left(1\right)}^{0}I_{1}^{i}-x_{\left(1\right)}^{i}\left(\Phi-I_{1}\right)\right]\\&-\frac{1}{2}\partial_{i}\partial_{j}\Phi\left(x_{\left(1\right)}^{i}+n^{i}x_{\left(1\right)}^{0}\right)\left(x_{\left(1\right)}^{j}+n^{j}x_{\left(1\right)}^{0}\right)-\frac{1}{2}x_{\left(1\right)}^{0}\left(4x_{\left(1\right)}^{0}n^{i}\partial_{i}\Phi'+\left(n^{i}x_{\left(1\right)}^{0}+2x_{\left(1\right)}^{i}\right)\partial_{i}A'\right).
    }
In (\ref{aomega_1})-(\ref{aomega_3}), all quantities are evaluated
at the background emission point at
$\bar{x}^{i}(\bar{\lambda}_e)\equiv (\lambda_e-\lambda_o)n^i$.

\section{Temperature anisotropy up to the third-order}

\subsection{Relation with primordial perturbations}


Having derived the perturbed photon energy in terms of the metric
perturbations at emission surface, the next goal is to relate the
temperature anisotropy to the primordial curvature perturbation
$\zeta$, which encodes the information in the very early universe
and is the most frequently used variable in evaluating the
primordial non-Gaussianities in the literature. It is well-known
that on large scales and for adiabatic perturbation, there is a
non-perturbative and conserved quantity which can be identified as
non-linear curvature perturbation in uniform-density slices
\cite{Salopek:1990jq,Kolb:2004jg,Lyth:2004gb,Rigopoulos:2003ak},
defined as{\footnote{The existence of non-perturbative conserved
perturbation has also been derived by using covariant formalism
\cite{Langlois:2005ii,Langlois:2005qp}, where the corresponding
quantity is defined as a co-vector: $\zeta_a = \partial_a \alpha -
\frac{\dot{\alpha}}{\dot{\rho}}\partial_a\rho$,  $\alpha\equiv
\frac{1}{3}\int d\tau \nabla_au^a$ is the local expansion. See
\cite{Langlois:2010vx} for a recent review.}}
    \eq{{\label{zeta_def}}
        \zeta \equiv -\Psi
        +\frac{1}{3}\int_{\bar{\rho}}^{\rho}\frac{d\tilde{\rho}}{3(\tilde{\rho}+\tilde{p})}.
    }
Conserved and gauge-invariant $\zeta$ beyond the linear theory has
also been constructed perturbatively in
\cite{Malik:2003mv,Lyth:2003im,Lyth:2005du}.
    The
conservation of $\zeta$ will allow us to relate the primordial era
when modes exit the horizon during inflation and the era when modes
re-enter the horizon, which is just the time of emission $\eta_e$
for our purpose. Our next task is to set the initial conditions for
$\Phi$ and $\Psi$ in the matter dominated era up to the third-order
in the primordial curvature perturbation $\zeta$.

In matter-dominated era ($p=0$),  (\ref{zeta_def}) can be integrated
to give
    \eq{
        \zeta = -\Psi +\frac{1}{3}\ln\frac{\rho_{m}}{\bar{\rho}_{m}} \,.
    }
On large scales, the matter density $\rho_{m}$ can be related to
metric perturbations through Einstein equation as (see Appendix
(\ref{app_Einstein_00})-(\ref{app_rho_m})):
    \eq{
        \frac{\rho_m}{\bar{\rho}_m} = e^{-2\Phi},
    }
which implies \cite{Bartolo:2005fp,D'Amico:2007iw}
\eq{{\label{zeta_final}}
        \zeta= -\Psi -\frac{2}{3}\Phi.
    }
(\ref{zeta_final}) is a non-perturbatively relation among $\zeta$
and $\Phi$, $\Psi$ on large scales during matter era.

At linear order, for fluid without anisotropic stress,
$\Phi_{(1)}=\Psi_{(1)}$, which gives the well-known relation $\zeta
= -\frac{5}{3}\Phi_{(1)}$. However, $\Phi\neq \Psi$ at non-linear
orders even for perfect fluid \cite{Acquaviva:2002ud,Bartolo:2003gh}
(see also
\cite{Bartolo:2003bz,Bartolo:2003jx,Boubekeur:2008kn,Christopherson:2009fp}
for the discussion of
 evolution of higher-order cosmological perturbations). From the
traceless part of $(i-j)$-component of Einstein equation and using
$(0-0)$ and $(0-i)$ components to express $\rho$ and $u_i$ in terms
of metric perturbations (see Appendix
(\ref{app_Einstein_00})-(\ref{app_Einstein_0i})), we are able to
write a non-perturbative constraint between $\Phi$ and $\Psi$ on
large scales during matter-dominated era
\cite{Bartolo:2005fp,D'Amico:2007iw},
    \ea{{\label{Phi_Psi_constr}}
        \partial^{4}\left(\Psi-\Phi\right) =&\frac{7}{2}\left(\partial^{2}\Phi\right)^{2}-\frac{3}{2}\left(\partial^{2}\Psi\right)^{2}+\frac{7}{6}\left(\partial_{i}\partial_{j}\Phi\right)^{2}-\frac{1}{2}\left(\partial_{i}\partial_{j}\Psi\right)^{2}+\frac{14}{3}\partial_{i}\Phi\partial_{i}\partial^{2}\Phi-2\partial_{i}\Psi\partial_{i}\partial^{2}\Psi\\&+\partial_{i}\partial_{j}\Phi\partial_{i}\partial_{j}\Psi+3\partial^{2}\Phi\partial^{2}\Psi+2\partial_{i}\Phi\partial_{i}\partial^{2}\Psi+2\partial_{i}\Psi\partial_{i}\partial^{2}\Phi.
    }
In deriving (\ref{Phi_Psi_constr}) we have neglected higher-order
spatial derivative terms since we are focusing on large scales. From
(\ref{Phi_Psi_constr}) $\Psi$ can be solved up to third-order in
$\Phi$ as
    \eq{{\label{Psi_Phi_sol}}
        \Psi
        =\Phi+\partial^{-4}\left[5\left(\partial^{2}\Phi\right)^{2}+\frac{5}{3}\left(\partial_{i}\partial_{j}\Phi\right)^{2}+\frac{20}{3}\partial_{i}\Phi\partial_{i}\partial^{2}\Phi\right].
    }
This is the generalization of the linear-order relation $\Psi=\Phi$
up to the third-order in $\Phi$. It is interesting to note the
third-order part of (\ref{Psi_Phi_sol}) exactly vanishes.

Combining (\ref{zeta_final}) and (\ref{Psi_Phi_sol}), it is now
straightforward to solve
$\Phi=\Phi[\zeta]=\Phi_{(1)}+\Phi_{(2)}+\Phi_{(3)}+\cdots$
perturbatively to give
    \begin{eqnarray}
        \Phi_{\left(1\right)} & = & -\frac{3}{5}\zeta, {\label{Phi_1}}\\
        \Phi_{\left(2\right)} & = &
                    -\frac{9}{25}\partial^{-4}\left[3\left(\partial^{2}\zeta\right)^{2}+\left(\partial_{i}\partial_{j}\zeta\right)^{2}+4\partial_{i}\zeta\partial_{i}\partial^{2}\zeta\right],
                    {\label{Phi_2}}
                    \end{eqnarray}
and
    \ea{{\label{Phi_3}}
        \Phi_{\left(3\right)}=&-\frac{54}{125}\partial^{-4}\Big[\left(3\partial^{2}\zeta\partial^{-2}+\partial_{i}\partial_{j}\zeta\partial_{i}\partial_{j}\partial^{-4}+2\partial_{i}\zeta\partial_{i}\partial^{-2}+2\partial_{i}\partial^{2}\zeta\partial_{i}\partial^{-4}\right)\\&\qquad\qquad\quad\times\left(3\left(\partial^{2}\zeta\right)^{2}+\left(\partial_{i}\partial_{j}\zeta\right)^{2}+4\partial_{i}\zeta\partial_{i}\partial^{2}\zeta\right)\Big] .
    }
    (\ref{Phi_1})-(\ref{Phi_3}) give the large-scale initial condition for
    $\Phi$ during the matter era, in terms of the conserved primordial curvature
    perturbation.
In the above $\partial^{-2}$ etc. can be understood in momentum
space. From (\ref{Phi_1})-(\ref{Phi_3}) and (\ref{Psi_Phi_sol}) the
corresponding initial condition $\Psi=\Psi[\zeta]$ up to the
third-order in $\zeta$ can also be easily get.

\subsection{Non-linear temperature anisotropy}

In the last part of this note, we will relate the observed
temperature anisotropy $\frac{\Delta T}{T}$ to the primordial
curvature perturbation $\zeta$, on large scales. To this end, we
also need the intrinsic temperature at the emission surface in terms
of metric perturbations. A fully treatment involves dynamics during
recombination
\cite{Bartolo:2006cu,Bartolo:2006fj,Nitta:2009jp,Pitrou:2008ak,Pitrou:2008hy,Pitrou:2008ut,Pitrou:2010sn,Beneke:2010eg}.
Here in this note, we take the large-scale non-perturbative
expression found in \cite{Bartolo:2005fp}, where in matter dominated
era with adiabatic assumption: $T_e = \bar{T}_e
e^{-\frac{2}{3}\Phi}$. Thus the large-scale temperature anisotropy
up to the third-order in $\Phi$ is given by $\frac{\Delta T}{T} =
\lrb{\frac{\Delta T}{T}}_{(1)}+\lrb{\frac{\Delta
T}{T}}_{(2)}+\lrb{\frac{\Delta T}{T}}_{(3)}+\cdots $ with
    \ea{
        \left(\frac{\Delta T}{T}\right)_{\left(1\right)} &=
        \frac{\Phi}{3}-I_{1},\\
        \left(\frac{\Delta T}{T}\right)_{\left(2\right)}&=\frac{\Phi^{2}}{18}+\frac{1}{3}\partial_{i}\Phi\left(n^{i}x_{\left(1\right)}^{0}+x_{\left(1\right)}^{i}\right)-\frac{\Phi
        I_{1}}{3}-I_{2}+x_{\left(1\right)}^{0}A',
    }
and
    \ea{
        \left(\frac{\Delta T}{T}\right)_{\left(3\right)}=&\frac{\Phi^{3}}{162}-\frac{\Phi^{2}I_{1}}{18}+x_{\left(2\right)}^{0}A'+\frac{1}{3}\Phi\left(x_{\left(1\right)}^{0}A'-I_{2}\right)-I_{3}+\frac{1}{2}x_{\left(1\right)}^{0}\left[\partial_{i}A'\left(n^{i}x_{\left(1\right)}^{0}+2x_{\left(1\right)}^{i}\right)-2I_{1}A'+x_{\left(1\right)}^{0}A''\right]\\&+\partial_{i}\Phi\left[\frac{1}{3}\left(x_{\left(2\right)}^{i}+x_{\left(1\right)}^{0}I_{1}^{i}\right)+\frac{1}{9}n^{i}\left(3x_{\left(2\right)}^{0}+x_{\left(1\right)}^{0}\left(\Phi+6A-18x_{\left(1\right)}^{0}-6I_{1}\right)\right)+x_{\left(1\right)}^{i}\left(\Phi-3I_{1}\right)\right]\\&+\frac{1}{6}\partial_{i}\partial_{j}\Phi\left(x_{\left(1\right)}^{i}+n^{i}x_{\left(1\right)}^{0}\right)\left(x_{\left(1\right)}^{j}+n^{j}x_{\left(1\right)}^{0}\right)+2n^{i}\partial_{i}\Phi'\left(x_{\left(1\right)}^{0}\right)^{2}.
    }

In the following, in order to further relate $\frac{\Delta T}{T}$ to
$\zeta$, we take the SW contribution \cite{Bartolo:2005fp} where we
neglect ISW and lensing contributions: $\frac{\Delta T}{T} =
\frac{\Phi}{3}+\frac{\Phi^{2}}{18}+\frac{\Phi^{3}}{162}+\cdots
\simeq e^{\frac{\Phi}{3}}-1$. Using (\ref{Phi_1})-(\ref{Phi_3}), in
momentum space, the non-linear mapping from $\zeta$ to $\frac{\Delta
T}{T}$ up to the third-order in $\zeta$ can be easily get. At linear
order we find the familiar relation  $\lrb{\frac{\Delta
T}{T}}_{(1)}= -\frac{1}{5}\zeta$, at the second-order and
third-order we find
\begin{eqnarray}
\left(\frac{\Delta T}{T}\right)_{\left(2\right)}\left(\bm{k}\right) & = & \frac{1}{2}\int\frac{d^{3}p_{1}d^{3}p_{2}}{\left(2\pi\right)^{3}}\delta^{3}\left(\bm{k}-\bm{p}_{1}-\bm{p}_{2}\right)\beta\left(k;p_{1},p_{2}\right)\zeta_{\bm{p}_{1}}\zeta_{\bm{p}_{2}}, \label{DeltaT2}\\
\left(\frac{\Delta T}{T}\right)_{\left(3\right)}\left(\bm{k}\right)
& = &
\frac{1}{3!}\int\frac{d^{3}p_{1}d^{3}p_{2}d^{3}p_{3}}{\left(2\pi\right)^{6}}\delta^{3}\left(\bm{k}-\bm{p}_{1}-\bm{p}_{2}-\bm{p}_{3}\right)\gamma\left(\bm{k};\bm{p}_{1},\bm{p}_{2},\bm{p}_{3}\right)\zeta_{\bm{p}_{1}}\zeta_{\bm{p}_{2}}\zeta_{\bm{p}_{3}},
\label{DeltaT3}\end{eqnarray} where the kernel
\begin{eqnarray}
\beta\left(k;p_{1},p_{2}\right) & = & -\frac{1}{50}+\frac{9\left(p_{1}^{2}-p_{2}^{2}\right)^{2}}{50k^{4}}-\frac{3\left(p_{1}^{2}+p_{2}^{2}\right)}{25k^{2}}, \label{beta_def}\\
\gamma\left(\bm{k};\bm{p}_{1},\bm{p}_{2},\bm{p}_{3}\right) & = &
-\frac{1}{125}+\left[\left(1-g\left(\bm{p}_{1},\bm{p}_{23}\right)\right)g\left(\bm{p}_{2},\bm{p}_{3}\right)+2\textrm{
cyclic}\right], \label{gamma_def}\end{eqnarray} with
$\bm{p}_{ij}=\bm{p}_i+\bm{p}_j$ and
    \eq{
        g\left(\bm{p},\bm{q}\right)=\frac{3}{250}\left[1+2\frac{p^{2}+q^{2}}{\left(\bm{p}+\bm{q}\right)^{2}}-3\frac{\left(p^{2}-q^{2}\right)^{2}}{\left(\bm{p}+\bm{q}\right)^{4}}\right].
    }
It is interesting to note that $\beta \rightarrow 1/25$ and
$g\rightarrow 0$ when $\bm{p}\rightarrow 0$ or $\bm{q}\rightarrow
0$, which implies that the non-linear mapping
(\ref{DeltaT2})-(\ref{DeltaT3}) would not contributes significantly
in the so-called ``squeezed" momenta configurations of
non-Gaussianity
\cite{Bartolo:2005fp,D'Amico:2007iw,Boubekeur:2009uk}.

Using (\ref{app_B_zeta}) and (\ref{app_T_zeta}), the bispectrum and
trispectrum for $\frac{\Delta T}{T}$ can be read as
    \eq{{\label{bispectrum}}
        B\left(k_{1},k_{2},k_{3}\right)=\alpha^{3}B_{\zeta}\left(k_{1},k_{2},k_{3}\right)+\left(\alpha^{2}P_{\zeta}\left(k_{1}\right)P_{\zeta}\left(k_{2}\right)\beta\left(k_{3};k_{1},k_{2}\right)+2\textrm{ cyclic}\right)
    }
and
    \ea{{\label{trispectrum}}
        T\left(\bm{k}_{1,}\bm{k}_{2},\bm{k}_{3},\bm{k}_{4}\right)
        =&\alpha^{4}T_{\zeta}\left(\bm{k}_{1,}\bm{k}_{2},\bm{k}_{3},\bm{k}_{4}\right)\\&+\alpha^{3}B_{\zeta}\left(k_{1,}k_{2},k_{12}\right)P_{\zeta}\left(k_{3}\right)\beta\left(k_{4};k_{12},k_{3}\right)+11\textrm{perms}\\&+\alpha^{3}P_{\zeta}\left(k_{1}\right)P_{\zeta}\left(k_{2}\right)P_{\zeta}\left(k_{3}\right)\gamma\left(\bm{k}_{4};-\bm{k}_{1},-\bm{k}_{2},-\bm{k}_{3}\right)+3\textrm{perms}\\&+\alpha^{2}P_{\zeta}\left(k_{1}\right)P_{\zeta}\left(k_{2}\right)P_{\zeta}\left(k_{13}\right)\beta\left(k_{3};k_{1},k_{13}\right)\beta\left(k_{4};k_{2},k_{13}\right)+11\textrm{perms}.
    }
Here $B_{\zeta}$ and $T_{\zeta}$ are primordial bispectrum and
trispectrum for the curvature perturbation $\zeta$ respectively.

In \cite{Komatsu:2001rj,Komatsu:2003fd}, the non-Gaussianities are
conventionally characterized by non-linear relation of the Bardeen
potential{\footnote{Here the Bardeen potential $\Phi$ should not be
confused with the metric perturbation $\Phi$ in (\ref{pert_metric}).
Actually at linear order, $\Phi_{\textrm{L}} = -\Phi_{(1)}$.}}
    \eq{{\label{f_g_NL_def}}
        \Phi = \Phi_{\textrm{L}} +
        f_{\textrm{NL}}\ast\Phi^2_{\textrm{L}} +
        g_{\textrm{NL}}\ast\Phi^3_{\textrm{L}},
    }
where $\Phi_{\textrm{L}}$ is the Gaussian part of $\Phi$ and
$f_{\textrm{NL}}$ and $g_{\textrm{NL}}$ are the so-called non-linear
parameters, ``$\ast$" denotes possible integration in momentum
space{\footnote{For example, $f_{\textrm{NL}}\ast\Phi^2_{\textrm{L}}
\equiv
\int\frac{d^{3}p_{1}d^{3}p_{2}}{\left(2\pi\right)^{3}}\delta^{3}\left(\bm{k}-\bm{p}_{1}-\bm{p}_{2}\right)f_{\textrm{NL}}\left(k;p_{1},p_{2}\right)\Phi_{\textrm{L}}\left(\bm{p}_{1}\right)\Phi_{\textrm{L}}\left(\bm{p}_{2}\right)$.}}.
To make contact with  previous analysis and conventions in the
literature, in the following we use the linear-order relations
during matter-dominated era: $\frac{\Delta T}{T}\equiv
-\frac{1}{3}\Phi$ and
$\frac{1}{3}\Phi_{\textrm{L}}=\frac{1}{5}\zeta_{\textrm{L}}$, and
make the ansatz for primordial non-Gaussianity: $\zeta =
\zeta_{\textrm{L}} + \frac{3}{5}f^{\zeta}_{\textrm{NL}}\ast
\zeta_{\textrm{L}}^2 + \frac{9}{25}g^{\zeta}_{\textrm{NL}}\ast
\zeta_{\textrm{L}}^3$. After some manipulations, the non-linear
parameters defined in (\ref{f_g_NL_def}) can be calculated as
    \eq{{\label{f_NL_final}}
        f_{\textrm{NL}}
        \left(k;p_{1},p_{2}\right)=f_{\textrm{NL}}^{\zeta}\left(k;p_{1},p_{2}\right)-\frac{25}{3}\beta\left(k;p_{1},p_{2}\right),
    }
and
    \ea{{\label{g_NL_final}}
        g_{\textrm{NL}}\left(\bm{k};\bm{p}_{1},\bm{p}_{2},\bm{p}_{3}\right)= &g_{\textrm{NL}}^{\zeta}\left(\bm{k};\bm{p}_{1},\bm{p}_{2},\bm{p}_{3}\right)\\
        &-\frac{50}{9}\left[\beta\left(k;p_{1},\left|\bm{k}-\bm{p}_{1}\right|\right)f_{\textrm{NL}}^{\zeta}\left(\left|\bm{k}-\bm{p}_{1}\right|,p_{2},p_{3}\right)+2\textrm{
        cyclic}\right]-\frac{125}{9}\gamma\left(\bm{k};\bm{p}_{1},\bm{p}_{2},\bm{p}_{3}\right).
    }
where $f_{\textrm{NL}}^{\zeta}$ and $g_{\textrm{NL}}^{\zeta}$ are
 non-linear parameters for primordial curvature perturbation
 $\zeta$, the functions $\beta$ and $\gamma$ are given in (\ref{beta_def}) and (\ref{gamma_def}).
In (\ref{f_NL_final})-(\ref{g_NL_final}), different contributions to
the finally observed non-Gaussianity from primordial epoch and
 from non-linearity between $\zeta$ and $\frac{\Delta T}{T}$ are transparent.

\section{Conclusion}

In this note, following the approach developed in
\cite{Pyne:1995bs,Mollerach:1997up,Matarrese:1997ay}, the general
expression for the observed CMB anisotropy up to the third-order is
calculated \cite{D'Amico:2007iw}. In the Sachs-Wolfe limit, we
derive the non-linear relation between the observed  anisotropy and
the conserved primordial curvature perturbation $\zeta$, up to the
third-order in $\zeta$. (\ref{DeltaT2})-(\ref{DeltaT3}) can be
viewed as non-linear generalization of familiar linear relation
$\Delta T/T= -\zeta/5$. Our formalism is valid for large-scale
anisotropies, which re-enter the horizon after decoupling. The
results (\ref{bispectrum})-(\ref{trispectrum}) clearly show the
different contributions to the observed non-Gaussianity from
primordial non-Gaussianities in $\zeta$ and non-linear mapping from
$\zeta$ to $\frac{\Delta T}{T}$ due to gravitational perturbations.
We also derive the non-linear parameters $f_{\textrm{NL}}$ and
$g_{\textrm{NL}}$ (eq.(\ref{f_NL_final})-(\ref{g_NL_final})), which
enter the theoretical predictions for the angular bispectrum and
trispectrum of CMB respectively.

We do not expect the non-linear mapping in the SW limit
(\ref{DeltaT2})-(\ref{DeltaT3}) would give a major contribution to
the observed non-Gaussianities, especially comparing scenarios where
large ``primordial" non-Gaussianities can be generated (see
\cite{Koyama:2010xj} for a recent review and references therein).
However, the other secondary anisotropies, especially the
correlation between lensing and ISW effect which we do not discuss
in this note, is expected to give contribution to the final
non-Gaussianity
\cite{Seljak:1998nu,Goldberg:1999xm,Serra:2008wc,Hanson:2009kg,Mangilli:2009dr,Creminelli:2004pv,Cooray:2002ee}.
Another interesting issue is that, if an enhancement of
non-linearity between $\frac{\Delta T}{T}$ and $\zeta$ is possible,
like the enhancement of primordial non-Gaussianity by small $c_s$
etc. in some inflationary scenarios. We wish to come back to these
subjects in future investigations.


\acknowledgments

I appreciate Yi-Fu Cai, Tao-Tao Qiu and Yi Wang for useful
discussion and comments. I am grateful to Prof. Miao Li for
  consistent
encouragement and support. This work was supported by the NSFC grant
No.10535060/A050207, a NSFC group grant No.10821504 and Ministry of
Science and Technology 973 program under grant No.2007CB815401.


\appendix

\section{Non-Gaussianities from non-linear mapping}

Non-Gaussian variables can be get from non-linear mapping from
Gaussian/non-Gaussian variables. The $\delta N$-formalism is a
non-linear mapping from inflaton fluctuation $\delta\phi$ to the
curvature perturbation $\zeta$ on super-Hubble scales. In general,
the non-linear mapping in real space from a single variable $Q$ to
$\zeta$ comes from three types: local products (e.g. $\delta
N$-formalism), products of local derivative terms, products of
non-local terms. In all cases, the mapping can be written in fourier
space as:
    \eq{{\label{app_nl_map}}
        \zeta_{\bm{k}} = \alpha\, Q_{\bm{k}}+\frac{1}{2!}\int\widetilde{dp_{1}}\widetilde{dp_{2}}\,\tilde{\beta}_{k;p_{1},p_{2}}Q_{\bm{p}_{1}}Q_{\bm{p}_{2}}+\frac{1}{3!}\int\widetilde{dp_{1}}\widetilde{dp_{2}}\widetilde{dp_{3}}\,\tilde{\gamma}_{\bm{k};\bm{p}_{1},\bm{p}_{2},\bm{p}_{3}}Q_{\bm{p}_{1}}Q_{\bm{p}_{2}}Q_{\bm{p}_{3}}+\cdots,
    }
with $\widetilde{dp}\equiv\frac{d^{3}p}{\left(2\pi\right)^{3}}$,
$\alpha$ is a $\bm{k}$-independent number and
\begin{eqnarray}
\tilde{\beta}_{\bm{k};\bm{p}_{1},\bm{p}_{2}} & \equiv & \left(2\pi\right)^{3}\delta\left(\bm{k}-\bm{p}_{1}-\bm{p}_{2}\right)\beta\left(k;p_{1},p_{2}\right),\\
\tilde{\gamma}_{\bm{k};\bm{p}_{1},\bm{p}_{2},\bm{p}_{3}} & \equiv &
\left(2\pi\right)^{3}\delta\left(\bm{k}-\bm{p}_{1}-\bm{p}_{2}-\bm{p}_{3}\right)\gamma\left(\bm{k};\bm{p}_{1},\bm{p}_{2},\bm{p}_{3}\right),\end{eqnarray}
where $\beta\left(k;p_{1},p_{2}\right)$ and
$\gamma\left(\bm{k};\bm{p}_{1},\bm{p}_{2},\bm{p}_{3}\right)$ are
normal functions. It is useful to note that $\beta$ and $\gamma$ are
symmetric with respect to all $\bm{p}_i$'s.

It is more convenient to calculate the correlation functions of
$\hat{\zeta}\equiv \zeta-\left\langle \zeta\right\rangle $ since
$\langle \hat{\zeta}\rangle=0 $. Straightforward calculation gives
(subscript ``c" denotes connected contribution):
    \ea{
        \left\langle
        \hat{\zeta}_{\bm{k}_{1}}\hat{\zeta}_{\bm{k}_{2}}\hat{\zeta}_{\bm{k}_{3}}\right\rangle &=
        \left(2\pi\right)^{3}\delta^{3}\left(\sum_{i=1}^3\bm{k}_{i}\right)B_{\zeta}\left(k_{1},k_{2},k_{3}\right),\\
        \left\langle
        \hat{\zeta}_{\bm{k}_{1}}\hat{\zeta}_{\bm{k}_{2}}\hat{\zeta}_{\bm{k}_{3}}\hat{\zeta}_{\bm{k}_{4}}\right\rangle_c
        &=\left(2\pi\right)^{3}\delta\left(\sum_{i=1}^4\bm{k}_{i}\right)T_{\zeta}\left(\bm{k}_{1,}\bm{k}_{2},\bm{k}_{3},\bm{k}_{4}\right),
    }
with the leading contributions:
    \eq{{\label{app_B_zeta}}
B_{\zeta}\left(k_{1},k_{2},k_{3}\right)=\alpha^{3}B_{Q}\left(k_{1},k_{2},k_{3}\right)+\left(\alpha^{2}P_{Q}\left(k_{1}\right)P_{Q}\left(k_{2}\right)\beta\left(k_{3};k_{1},k_{2}\right)+\textrm{cyclic}\right),
} and
    \ea{{\label{app_T_zeta}}
        T_{\zeta}\left(\bm{k}_{1,}\bm{k}_{2},\bm{k}_{3},\bm{k}_{4}\right)
        =&\alpha^{4}T_{Q}\left(\bm{k}_{1,}\bm{k}_{2},\bm{k}_{3},\bm{k}_{4}\right)\\&+\alpha^{3}B_{Q}\left(k_{1,}k_{2},k_{12}\right)P_{Q}\left(k_{3}\right)\beta\left(k_{4};k_{12},k_{3}\right)+11\textrm{perms}\\&+\alpha^{3}P_{Q}\left(k_{1}\right)P_{Q}\left(k_{2}\right)P_{Q}\left(k_{3}\right)\gamma\left(\bm{k}_{4};-\bm{k}_{1},-\bm{k}_{2},-\bm{k}_{3}\right)+3\textrm{perms}\\&+\alpha^{2}P_{Q}\left(k_{1}\right)P_{Q}\left(k_{2}\right)P_{Q}\left(k_{13}\right)\beta\left(k_{3};k_{1},k_{13}\right)\beta\left(k_{4};k_{2},k_{13}\right)+11\textrm{perms}    }
with $k_{ij}\equiv |\bm{k}_i +\bm{k}_j|$. In  the above $P_Q$ is the
power spectrum of $Q$, $B_{Q}$ and $T_{Q}$ are ``intrinsic"
bispectrum and trispectrum for $Q$ respectively which vanish if $Q$
is purely Gaussian.

\begin{figure}[h]
    \centering
    \includegraphics[width=3.5cm]{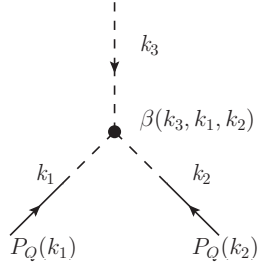}
    \caption{Diagrammatic representation of $P(k_1)P(k_2)\beta(k_3,k_1,k_2)$ in (\ref{app_B_zeta}).}
    \label{fig:B_zeta}
\end{figure}

\begin{figure}[h]
    \centering
    \begin{minipage}{0.28\textwidth}
    \includegraphics[width=4cm]{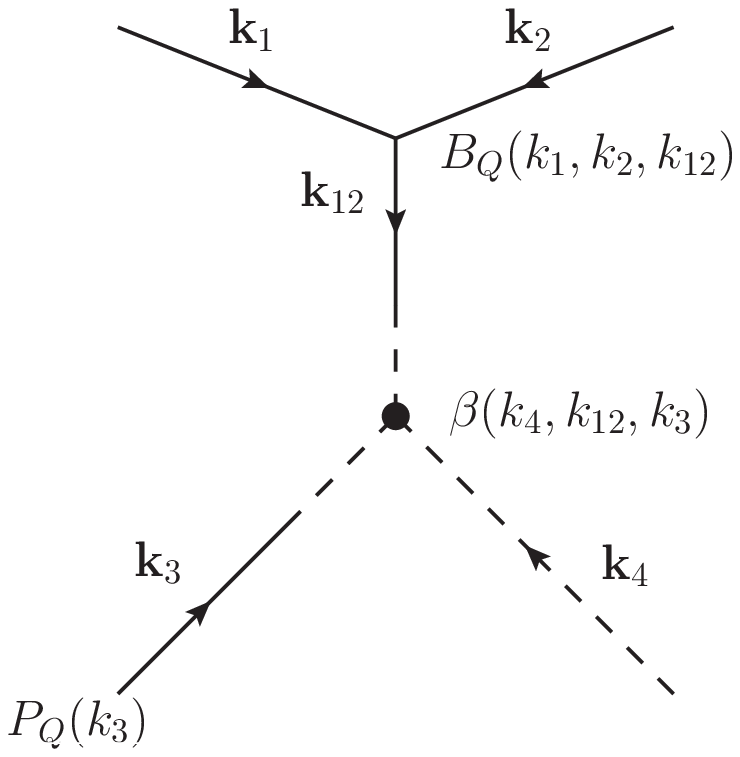}
    \end{minipage}
    \begin{minipage}{0.3\textwidth}
    \includegraphics[width=4.5cm]{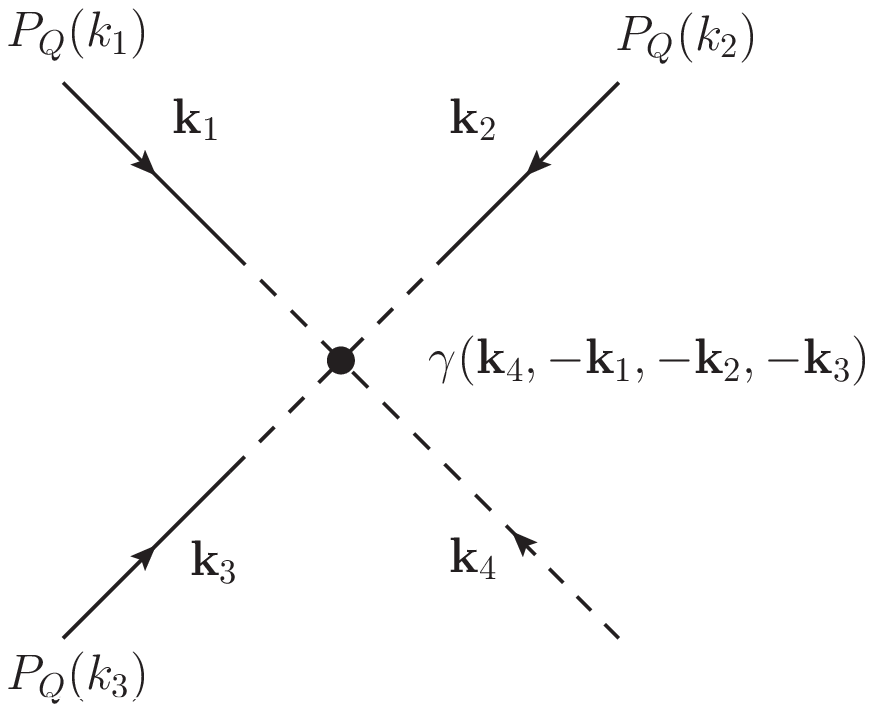}
    \end{minipage}
    \begin{minipage}{0.35\textwidth}
    \includegraphics[width=6cm]{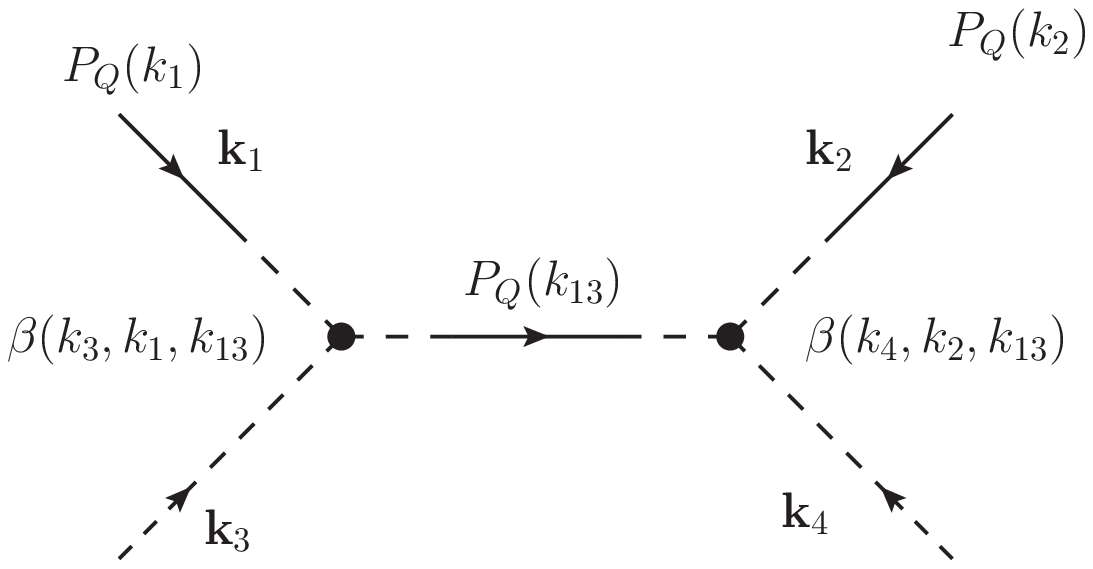}
    \end{minipage}
    \caption{Diagrammatic representation of the second, third and fourth line in (\ref{app_T_zeta}).}
    \label{fig:T_zeta}
\end{figure}

In \cite{Byrnes:2007tm}, based on the $\delta N$-formalism,
Feynman-type diagrams were introduced to represent various
contributions to non-Gaussianity from the non-linear mapping from
$\delta\phi$ to $\zeta$ on large scales. In general this can be
generalized straightforwardly for the non-linear mapping
(\ref{app_nl_map}), as we show in fig.\ref{fig:B_zeta} and
fig.\ref{fig:T_zeta}.

\section{Perturbed metric and related quantities}

For the perturbed conformal metric $g_{\mu\nu }$ defined in
(\ref{pert_metric}), the components of Christoffel symbol are
    \eq{
        \Gamma_{00}^{0}=\frac{e^{2\Phi}}{N^{2}}\left[\Phi'+e^{2\left(\Psi-\Phi\right)}e^{-\gamma_{ij}}\sigma_{j}\left(\sigma_{i}'+e^{2\Phi}\partial_{i}\Phi\right)\right],
    }
    \eq{
        \Gamma_{0i}^{0}= \Gamma_{i0}^{0} = \frac{e^{2\Phi}}{N^{2}}\left\{
        \partial_{i}\Phi+\frac{1}{2}e^{2\left(\Psi-\Phi\right)}e^{-\gamma_{jm}}\sigma_{m}\left[e^{-2\Psi}\left(-2\Psi'e^{\gamma_{ij}}+\left(e^{\gamma_{ij}}\right)'\right)+\partial_{i}\sigma_{j}-\partial_{j}\sigma_{i}\right]\right\},
    }
    \ea{
        \Gamma_{ij}^{0} =&\frac{e^{-2\Phi}}{2N^{2}}\Big\{-\partial_{i}\sigma_{j}-\partial_{j}\sigma_{i}-2\Psi'e^{-2\Psi}e^{\gamma_{ij}}+e^{-2\Psi}\left(e^{\gamma_{ij}}\right)'\\&-e^{-\gamma_{kl}}\sigma_{l}\left[\left(2\partial_{i}\Psi e^{\gamma_{jk}}-\partial_{i}e^{\gamma_{jk}}\right)+\left(2\partial_{j}\Psi e^{\gamma_{ki}}-\partial_{j}e^{\gamma_{ki}}\right)-\left(2\partial_{k}\Psi e^{\gamma_{ij}}-\partial_{k}e^{\gamma_{ij}}\right)\right]\Big\}.
    }
    \eq{
        \Gamma_{00}^{i} =
        -\frac{e^{2\left(\Phi+\Psi\right)}}{N^{2}}\Phi'e^{-\gamma_{ij}}\sigma_{j}+e^{2\Psi}e^{-\gamma_{ik}}\left(\delta_{kj}-\frac{e^{2\Psi}\sigma_{k}e^{-\gamma_{jn}}\sigma_{n}}{N^{2}}\right)\left(\sigma_{j}'+e^{2\Phi}\partial_{j}\Phi\right),
    }
    \ea{
        & \Gamma_{0j}^{i}\equiv\Gamma_{j0}^{i} \\
        =&
        -\frac{e^{2\left(\Phi+\Psi\right)}}{N^{2}}e^{-\gamma_{ik}}\sigma_{k}\partial_{j}\Phi+\frac{1}{2}\left(e^{-\gamma_{ik}}-\frac{e^{2\Psi}}{N^{2}}\left(e^{-\gamma_{im}}\sigma_{m}\right)\left(e^{-\gamma_{kn}}\sigma_{n}\right)\right)\left[\left(-2\Psi'e^{\gamma_{jk}}+\left(e^{\gamma_{jk}}\right)'\right)+e^{2\Psi}\left(\partial_{j}\sigma_{k}-\partial_{k}\sigma_{j}\right)\right],
    }
    \ea{
        \Gamma_{jk}^{i} =&\frac{1}{2}\frac{e^{-\gamma_{il}}\sigma_{l}}{N^{2}}\left[e^{2\Psi}\left(\partial_{j}\sigma_{k}+\partial_{k}\sigma_{j}\right)+2\Psi'e^{\gamma_{jk}}-\left(e^{\gamma_{jk}}\right)'\right]\\&-\frac{1}{2}\left(e^{-\gamma_{il}}-\frac{e^{2\Psi}}{N^{2}}\left(e^{-\gamma_{im}}\sigma_{m}\right)\left(e^{-\gamma_{ln}}\sigma_{n}\right)\right)\left[\left(2\partial_{j}\Psi e^{\gamma_{kl}}-\partial_{j}e^{\gamma_{kl}}\right)+\left(2\partial_{k}\Psi e^{\gamma_{lj}}-\partial_{k}e^{\gamma_{lj}}\right)-\left(2\partial_{l}\Psi e^{\gamma_{jk}}-\partial_{l}e^{\gamma_{jk}}\right)\right].
    }
In the above
$N^2=e^{2\Phi}+e^{2\Psi}e^{-\gamma_{ij}}\sigma_{i}\sigma_{j}$, where
$N$ is the corresponding laspe function in ADM formalism.

For metric $ds^2 =
a^{2}\left(-e^{2\Phi}d\eta^{2}+e^{-2\Psi}dx^{i}dx^{i}\right)$, the
Christoffel connection is significantly simplified, with components:
    \ea{{\label{app_connection_ls}}
        \Gamma_{00}^{0} &= \Phi',\qquad
        \Gamma_{0i}^{0}=\partial_{i}\Phi,\qquad \Gamma_{ij}^{0} =
        -\Psi'e^{-2\left(\Phi+\Psi\right)}\delta_{ij},\qquad \Gamma_{00}^{i}=e^{2\left(\Phi+\Psi\right)}\partial_{i}\Phi,\\
        \Gamma_{0j}^{i}&=-\Psi'\delta_{ij},\qquad \Gamma_{jk}^{i}=-\left(\partial_{j}\Psi\delta_{ki}+\partial_{k}\Psi\delta_{ij}-\partial_{i}\Psi\delta_{jk}\right).
    }
 The corresponding components of Einstein tensor are
    \eq{
        G_{00} =
        3\left(\mathcal{H}-\Psi'\right)^{2}-e^{2(\Phi+\Psi)}\left(\left(\partial_{i}\Psi\right)^{2}-2\partial^{2}\Psi\right),
    }
    \eq{
        G_{0i} =
        2\left[\partial_{i}\Psi'+\partial_{i}\Phi\left(\mathcal{H}-\Psi'\right)\right],
    }
    \ea{
        G_{ij} =&\left\{ e^{-2(\Phi+\Psi)}\left[-2\mathcal{H}'-\left(\mathcal{H}-2\Phi'-3\Psi'\right)\left(\mathcal{H}-\Psi'\right)+2\Psi''\right]+\left(\partial_{i}\Phi\right)^{2}+\partial^{2}\left(\Phi-\Psi\right)\right\} \delta_{ij}\\&+\partial_{i}\partial_{j}\left(\Psi-\Phi\right)-\partial_{i}\Phi\partial_{j}\Phi+\partial_{i}\Psi\partial_{j}\Psi-\left(\partial_{i}\Phi\partial_{j}\Psi+\partial_{i}\Psi\partial_{j}\Phi\right).
    }
The above expressions are exact, which can be easily expanded to the
desired orders.

The $(00)$-component of Einstein equation $G_{\mu\nu} = T_{\mu\nu}$
gives,
    \eq{{\label{app_Einstein_00}}
        \frac{\rho_m}{\bar{\rho}_m} =
        \frac{3\left(\mathcal{H}-\Psi'\right)^{2}-e^{2(\Phi+\Psi)}\left(\left(\partial_{i}\Psi\right)^{2}-2\partial^{2}\Psi\right)}{1+a^{-2}e^{2\Psi}u_{i}u_{i}}\frac{e^{-2\Phi}}{3\mathcal{H}^{2}},
    }
where we used $\bar{\rho}_m=\frac{3\mathcal{H}^{2}}{a^{2}}$ on the
background level. From the $(0i)$-component of Einstein equation we
can solve
    \eq{{\label{app_Einstein_0i}}
        1+a^{-2}e^{2\Psi}u_{i}u_{i} =
        \frac{\left[3\mathcal{H}^{2}-e^{2(\Phi+\Psi)}\left(\left(\partial_{i}\Psi\right)^{2}-2\partial^{2}\Psi\right)\right]^{2}}{\left[3\mathcal{H}^{2}-e^{2(\Phi+\Psi)}\left(\left(\partial_{i}\Psi\right)^{2}-2\partial^{2}\Psi\right)\right]^{2}-4\mathcal{H}^{2}e^{2\left(\Phi+\Psi\right)}\left(\partial_{i}\Phi\right)^{2}}.
    }
where we have set $\Psi'=0$. Finally, (\ref{app_Einstein_00}) and
(\ref{app_Einstein_0i}) imply
    \eq{{\label{app_rho_m}}
        \frac{\rho_m}{\bar{\rho}_m} =
        e^{-2\Phi}\left[1-\frac{e^{2(\Phi+\Psi)}}{3\mathcal{H}^{2}}\left(\left(\partial_{i}\Psi\right)^{2}-2\partial^{2}\Psi\right)-\frac{4e^{2\left(\Phi+\Psi\right)}}{9\mathcal{H}^{2}}\frac{\left(\partial_{i}\Phi\right)^{2}}{1-\frac{e^{2(\Phi+\Psi)}}{3\mathcal{H}^{2}}\left(\left(\partial_{i}\Psi\right)^{2}-2\partial^{2}\Psi\right)}\right],
    }
which gives, on large scales,  ${\rho_m}/{\bar{\rho}_m} \simeq
e^{-2\Phi}$.



\end{document}